\documentclass[twocolumn,superscriptaddress,aps,pra,10pt]{revtex4-1}
\usepackage{amsmath}
\usepackage{graphicx}

\begin{document}

\title{Asymmetry of charge relaxation times in quantum dots:\\ The influence of degeneracy}

\author{Andreas Beckel}
\affiliation{Faculty of Physics and CENIDE, University of Duisburg-Essen, Lotharstr.~1, 47057 Duisburg, Germany}
\author{Annika Kurzmann}
\affiliation{Faculty of Physics and CENIDE, University of Duisburg-Essen, Lotharstr.~1, 47057 Duisburg, Germany}
\author{Martin Geller}
\affiliation{Faculty of Physics and CENIDE, University of Duisburg-Essen, Lotharstr.~1, 47057 Duisburg, Germany}
\author{Arne Ludwig}
\affiliation{Lehrstuhl f\"ur Angewandte Festk\"orperphysik, Ruhr-Universit\"at Bochum, Universit\"atstr.~150, 44780 Bochum, Germany}
\author{Andreas~D. Wieck}
\affiliation{Lehrstuhl f\"ur Angewandte Festk\"orperphysik, Ruhr-Universit\"at Bochum, Universit\"atstr.~150, 44780 Bochum, Germany}
\author{J\"urgen K\"onig}
\affiliation{Faculty of Physics and CENIDE, University of Duisburg-Essen, Lotharstr.~1, 47057 Duisburg, Germany}
\author{Axel Lorke}
\email{axel.lorke@uni-due.de}
\affiliation{Faculty of Physics and CENIDE, University of Duisburg-Essen, Lotharstr.~1, 47057 Duisburg, Germany}

\date{\today}

\begin{abstract}
Using time-resolved transconductance spectroscopy, we study the tunneling dynamics between a two-dimensional electron gas (2DEG) and self-assembled quantum dots (QDs), embedded in a field-effect transistor structure. 
We find that the tunneling of electrons from the 2DEG into the QDs is governed by a different time constant than the reverse process, i.e., tunneling from the QDs to the 2DEG. 
This asymmetry is a clear signature of Coulomb interaction and makes it possible to determine the degeneracy of the quantum dot orbitals even when the individual states cannot be resolved energetically because of inhomogeneous broadening.
Our experimental data can be qualitatively explained within a master-equation approach.
\end{abstract}

\maketitle

\section{Introduction}

Self-assembled InAs quantum dots are ideal model systems to study the energetic structure and dynamics of fully quantized few carrier systems \cite{Bimbergbuch,Petroff01,Reimann02}. When incorporated into a suitable diode or transistor structure, the coupling to a free electron or hole reservoir opens up new possibility for tuning the charge and energy of the dots \cite{Drexler94}. It also makes it possible to study the quantum mechanical properties in great detail. 

When investigating the non-equilibrium transport between a reservoir and the dot system, the charging and discharging dynamics are given by the tunneling matrix element, which gives access to, e.g.,  wave function mapping \cite{Vdovin00,Maltezopoulos03,Wibbelhoff05} and manipulation \cite{Rontani11,Lei10,Patane10}. 

As we will show in the following, also the multiplicity/degeneracy of the quantum dot states has a profound influence on the tunneling dynamics between the reservoir and the dots \cite{Reckermann10,Cockins10}. Starting from the observation that \emph{charging and discharging of the dots are governed by different relaxation times}, we develop a non-equilibrium transport model based on a master equation. The comparison between the model and the experimental data allows us to determine the details of the degeneracy of the electronic $p$-shell. These details are usually hidden by the unavoidable inhomogeneous ensemble broadening of the energy structure, but can be resolved by studying the charging and discharging dynamics. 

\begin{figure}[b]
	\includegraphics[width=0.9\columnwidth]{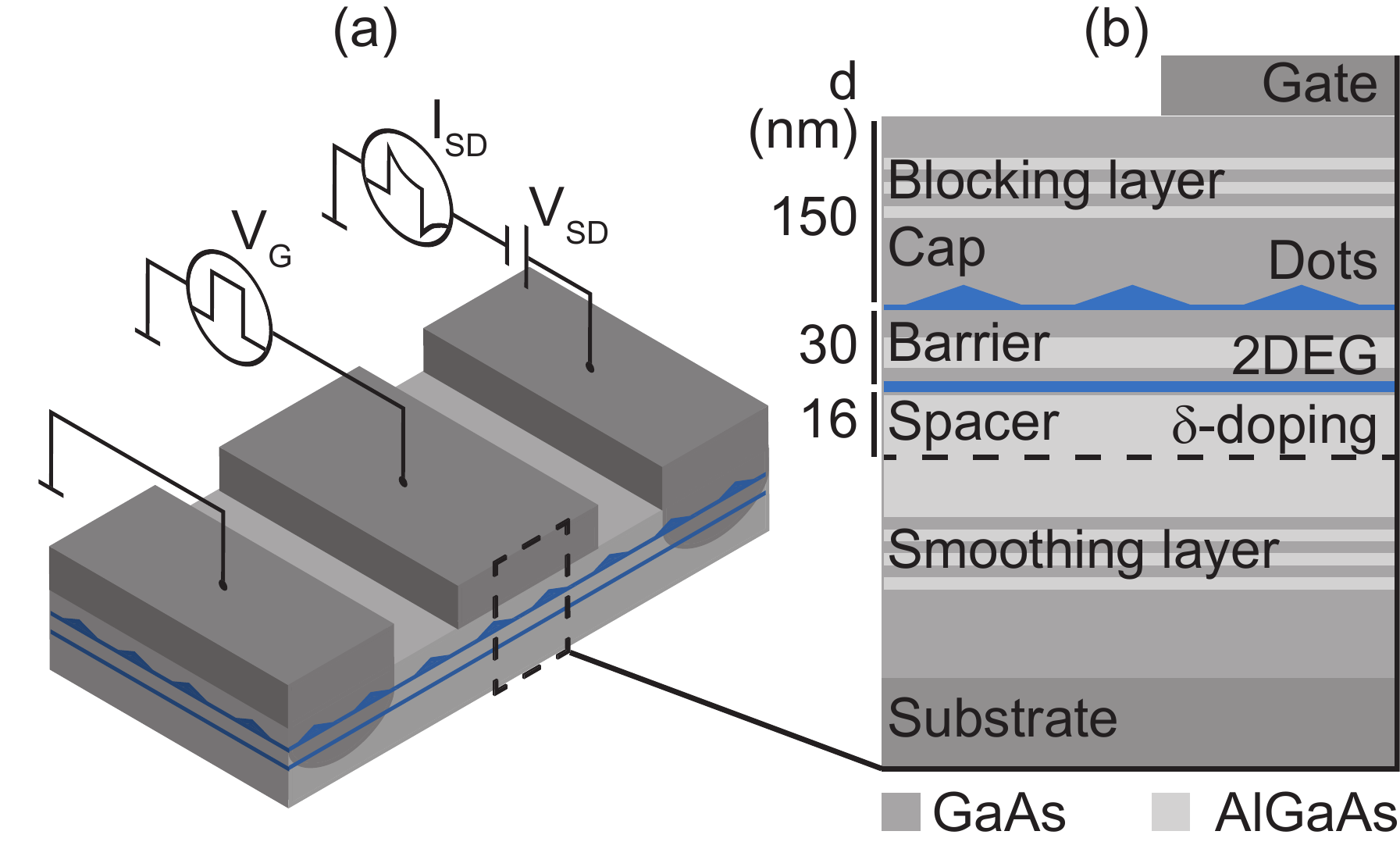}
	\caption{Device (a) and layer schematics (b) of the investigated sample: 
		A high electron mobility transistor, grown with an embedded layer of quantum dots, serving as a floating gate. 
		The tunneling current into the dots is monitored via a time-resolved conductance measurement of the 2DEG.
	}
	\label{Fig1}
\end{figure}

\section{Experiment}

The measurements are performed on an inverted AlGaAs/GaAs high electron mobility transistor structure as sketched in Fig. \ref{Fig1}(a) with an embedded layer of InAs quantum dots. 
The layer sequence, grown by molecular beam epitaxy, is schematically shown in Fig. \ref{Fig1}(b). 
The active part of the structure starts with a 300\,nm thick Al$_{0.34}$Ga$_{0.66}$As layer, a silicon $\delta$-doping sheet ($3\cdot 10^{12}$\,cm$^{-2}$) and an AlGaAs spacer layer of 16\,nm thickness. 
Subsequently, a 15\,nm thick GaAs layer, which contains a two-dimensional electron gas, a 10\,nm thick AlGaAs tunneling layer, a 5\,nm thick GaAs spacer layer and the InAs quantum dots are deposited. 
The dot formation takes place after evaporating the equivalent of 1.9 monolayers of InAs at 525$^\circ$C. 
This results in a dot density of $n_{\rm QD}\approx 8\cdot 10^9$ cm$^{-2}$. 
The dots are covered by 150\,nm of GaAs and a 116\,nm thick blocking layer of alternating AlAs/GaAs layers (3\,nm and 1\,nm, respectively). 
The structure is capped by a protective, 5\,nm thick GaAs film. 

Using standard lithographic techniques, the samples are patterned into a 60\,$\mu$m long and 50\,$\mu$m wide strip with source / drain contacts on either side. 
The central region is covered by a 50\,nm thick gold layer, which serves as a gate electrode. 
The application of a gate voltage $V_{\rm G}$ will shift the energetic position of the states in the quantum dots, which are embedded in the dielectric between the gate electrode and the two-dimensional electron gas (2DEG) \cite{Rus06PRB}. 
This way, the number of electrons per quantum dot can be adjusted between 0 and 6 \cite{Drexler94,Petroff01,Rus06PRB}. 
More specifically, each time the energy difference $\epsilon_m := E_{m} - E_{m-1}$ of the $m$-electron ground-state energies $E_m$ of the quantum dot is in resonance with the electro-chemical potential $\mu_{\rm F}$ of the two-dimensional reservoir, electrons can tunnel between 2DEG and quantum dots.

\begin{figure}[bt]
	\includegraphics[width=0.85\columnwidth]{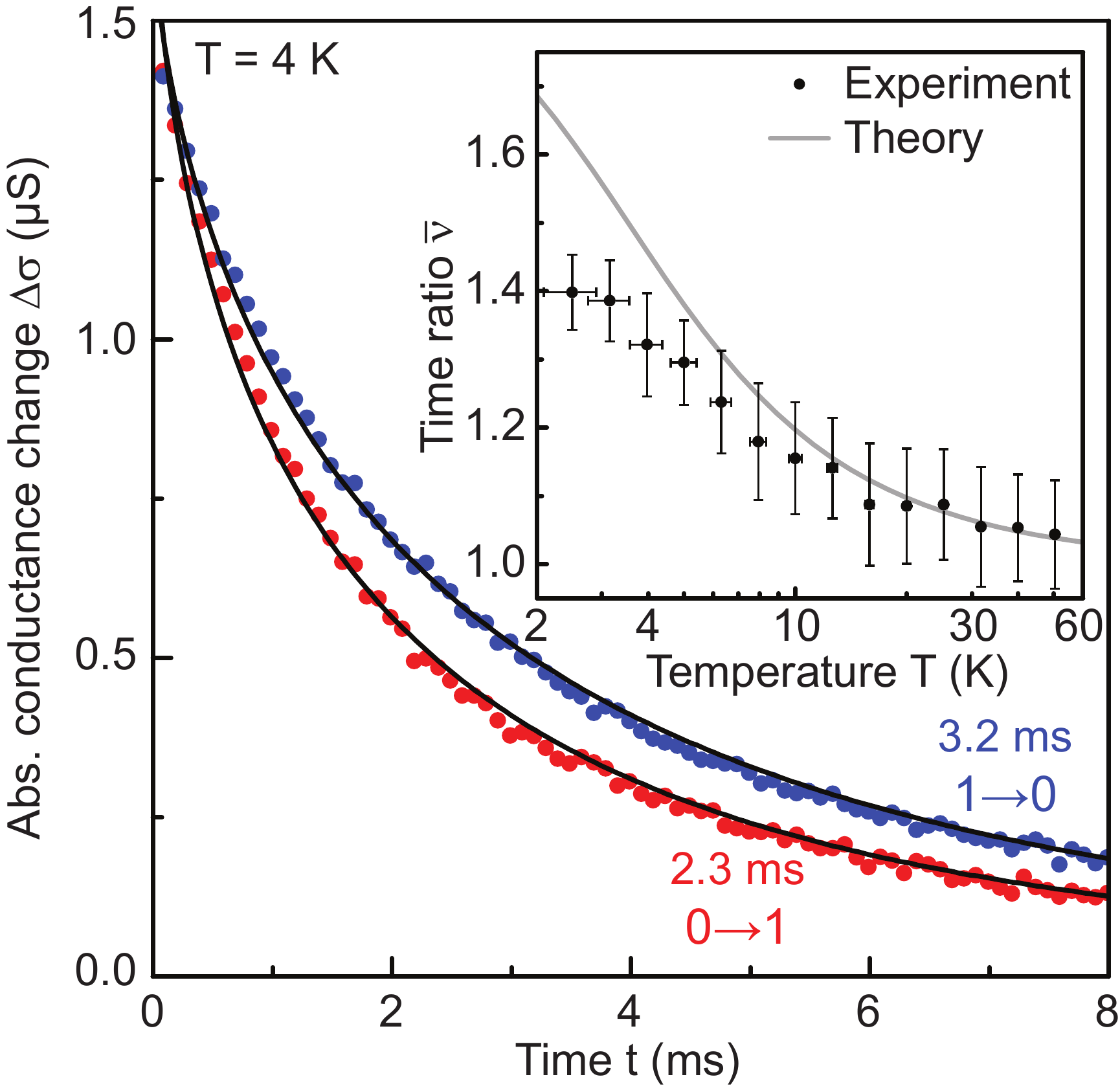}
	\caption{Conductance change $\Delta \sigma$ while charging ($0\rightarrow 1$) and discharging ($1\rightarrow 0$) the first electron ($V_{\rm G}=-0.67$~V). 
	The time constants are determined to $\tau_{0\rightarrow 1}$=2.3~ms and $\tau_{1\rightarrow 0}$=3.2~ms respectively by fitting a stretched exponential (solid lines) to the transients. 
	The inset shows the temperature dependence of the averaged time ratio $\bar{\nu}=(\nu_1+1/\nu_2)/2$ (see Eq.\ref{ratio}). 
	For sufficiently low temperatures we find a maximum ratio of $\approx 1.4$ while $\bar\nu\rightarrow 1$ for high temperatures. 
	The solid line shows the temperature dependence, calculated from a master equation (see text). 
	}
	\label{Fig2}
\end{figure}

To monitor the tunneling dynamics between the dot ensemble and the 2DEG, we use a recently developed transconductance spectroscopy technique \cite{Marquardt09,Marquardt11,Beckel12}. 
At a time $t=0$, a voltage pulse is applied to the gate, and the time-resolved response of the 2DEG conductivity $\sigma(t)$ is recorded. 
For a positive pulse (upward step in $V_{\rm G}$) the energy of the quantum dot states are shifted downward, so that electrons can tunnel from the 2DEG into unoccupied states in the dot layer. 
Therefore, an exponential decrease of $\sigma$ is observed, because mobile charges from the 2DEG will become localized when they are transferred into the dots. 
For the reverse process (when switching back to the original voltage), charges are transferred back out of the dots into the 2DEG, so that its conductance will increase again \cite{Rus06PRB,Marquardt09}.
In this way, the conductivity of the 2DEG is a direct measure of quantum-dot charge and the conductance traces as shown in Fig. \ref{Fig2} allow us to directly compare the charging with the discharging process. 

Taking the geometric distance between the 2DEG and the dot layer, $d_{\rm tunn}$, as well as the distance between the 2DEG and the gate, $d_{\rm tot}$, the energy shift $\Delta E$ caused by the voltage step $\Delta V_{\rm G}$ is easily calculated as $\Delta E =  e \frac{d_{\rm tunn}}{d_{\rm tot}} \Delta V_{\rm G} = \frac{e}{\lambda} \Delta V_{\rm G}$. 
Here we chose the simple but well established \cite{MedeirosRibeiro1997,Warburton98} energy conversion based on the geometric lever arm \cite{Note1} $\lambda = \frac{d_{\rm tot}}{d_{\rm tunn}}=7$. 

For small excitation voltages $\Delta V_{\rm G}$ this allows us to derive the density of states in the dot layer $D(E)$ from the measured total change in conductivity $\Delta \sigma=\left| \sigma(0)-\sigma(\infty)\right|$ from:
\begin{equation}
	\label{DOS}
	\frac{\Delta \sigma}{\Delta V_{\rm G}} = \frac{\Delta n e \mu}{\frac{\Delta E}{e \lambda}} = \lambda e^2 \mu \frac{\Delta n}{\Delta E} =  \lambda e^2 \mu D(E) , 
\end{equation}
where $\mu$ is the mobility of the 2DEG, and $\Delta n$ is the change in the 2DEG carrier density, caused by the tunneling electrons \cite{Note2}. 
Figure \ref{Fig3}(b) shows the thus obtained density of states in the dot layer. 
We observe two clearly distinct maxima, corresponding to the charging of the two $s$-states around $V_{\rm G} = -0.6$ V and a broader distribution, corresponding to the charging of the four $p$-states in the range between $-0.3$ and $0.3$ V. 
The peaks are broadened because of the size distribution of the self-assembled quantum dots. 
On samples with even better size homogeneity, the four $p$-states can also be clearly resolved \cite{Rus06PRB}. 

Turning to the dynamics of the tunneling process, we find a significant difference for the charging compared to the discharging process as shown in Fig.~\ref{Fig2}. 
Here, a small energy shift of $\Delta E\sim 1.4\,{\rm meV}$ in the chemical potential at a gate voltage $V_{\rm G} = -0.67\,{\rm V}$ allows a fraction of the dot ensemble \cite{Note3} to become charged ($0\rightarrow 1$) or discharged ($1\rightarrow 0$) with a single electron. 
From a stretched exponential fit \cite{Note4} we obtain time constants of $\tau_{0\rightarrow 1}=2.3$~ms and $\tau_{1\rightarrow 0}=3.2$~ms, which clearly differ from each other. 
As shown in the inset of Fig.~\ref{Fig2}, the ratio between the charging and discharging relaxation rate decreases with increasing temperature.
This raises two questions: 1) What is the physical origin of this asymmetry and 2) how can this asymmetry be used to gain insight in the internal structure of the quantum dots? 
To answer both questions we model the charge relaxation after the voltage pulse within a master-equation approach.

\section{Theory}

At first glance, an asymmetry between charging and discharging the dots may appear counter intuitive.
After all, Fermi's Golden Rule \cite{Dirac27,Fermi50} $\Gamma_{i \rightarrow f} = \frac{2 \pi}{\hbar} \left| \left< f \right|H'  \left| i \right>\right|^2 \rho_f$ for the transition rate from an initial state $i$ to a (fixed) final state $f$ is symmetric, $\Gamma_{i \rightarrow f} = \Gamma_{f \rightarrow i}\equiv \Gamma(\epsilon)$, as a consequence of the hermiticity of the tunneling Hamiltonian $H'$  and energy conservation which ensures that the (many-body) density of states for the initial and the final state are equal to each other (here practically given by the density of states of the 2DEG).
The dependence of $\Gamma(\epsilon)$ on the quantum-dot energy level $\epsilon$ reflects the energy dependence of the tunnel amplitudes (density of states of the 2DEG is practically energy independent).

We describe the charge dynamics by a master equation \cite{Beenakker} $\dot{p}_m = \sum_{m' \neq m} \Gamma_{m' \rightarrow m} p_{m'} - \sum_{m' \neq m} \Gamma_{m \rightarrow m'} p_m$ in terms of the quantum-dot charge $m$ (and its probability $p_m$), which contains only partial information of the initial and final many-body states. 
The 2DEG degrees of freedom and the $d_m$-fold (spin and/or orbital) degeneracy of the quantum-dot state with $m$ electrons are integrated out. 
Averaging the Fermi-Golden-Rule expression $\Gamma(\epsilon)$ over all initial and summing over all final states with the given quantum-dot charge yields the transition rates
\begin{eqnarray}
\label{Gm+}
	\Gamma_m^+ &\equiv& \Gamma_{m-1\rightarrow m} = k_{m-1\rightarrow m} \Gamma(\epsilon_m) f(\epsilon_m)
	\\
\label{Gm-}
	\Gamma_m^- &\equiv& \Gamma_{m\rightarrow m-1} = k_{m\rightarrow m-1} \Gamma(\epsilon_m) [1 - f(\epsilon_m)] \, ,
\end{eqnarray}
where the Fermi function $f(\epsilon)$ stems from the average over the 2DEG occupation. Only transitions with $m \leftrightarrow m- 1$ need to be taken into account, because the electron-electron interaction energy (Coulomb blockade) is much larger than both the thermal energy and the excitation energy induced by the voltage pulse.
The integer $k_{m\rightarrow m'}$ counts how many quantum-dot states with charge $m'$ can be reached from each of the states with charge $m$.
Due to selection rules, $k_{m\rightarrow m'}$ may be smaller than the degeneracy $d_{m'}$.
Nevertheless, their ratios are equal,
\begin{equation}
	\xi_m = \frac{k_{m-1\rightarrow m}}{k_{m\rightarrow m-1}} = \frac{d_{m}}{d_{m-1}} \, .
\end{equation}

Let us consider the $m$-th charge transition for an individual quantum dot.
From the master equations for the probabilities $p_{m-1}=1-p_{m}$, we obtain the kinetic equation for the average charge $N= \sum_m m p_m$,
\begin{equation}
 	\dot N(t) = m \Gamma_m^+ + (m-1) \Gamma_m^- - (\Gamma_m^+ + \Gamma_m^-) N(t)
\end{equation}
which is solved by
\begin{equation}
	\label{singleoccupation}
 	\Delta N(t) \equiv N(t) - N_{\rm eq} = \left( N_0 - N_{\rm eq} \right) \exp ( -t/\tau)
\end{equation}
with the relaxation time $\tau$ given by
\begin{equation}
\label{time}
	\frac{1}{\tau} = \tilde \Gamma_m \left[ 1 + (\xi_m -1) f(\epsilon_m) \right] \, ,
\end{equation}
where $\tilde \Gamma_m = k_{m\rightarrow m-1} \Gamma (\epsilon_m)$,
and the equilibrium occupation 
\begin{equation}
	\label{final}
 	N_{\rm eq} = m-1 + \frac{\xi_m f(\epsilon_m)}{1 + (\xi_m -1) f(\epsilon_m)}
	\, .
\end{equation}
We would like to mention that for a given (fixed) final state energy $\epsilon_m$, the relaxation time $\tau$ does {\it not} distinguish between charging and discharging, i.e., whether the initial charge $N_0$ was larger or smaller than $N_{\rm eq}$. 

Experimentally, on the other hand, the applied gate voltage pulse changes the energy $\epsilon_m$ of the quantum dot by a small amount $\Delta E$. For small voltage pulses, the energy dependence of the tunnel barrier can be neglected, i.e., $\Gamma(\epsilon) \approx \Gamma = const.$. However, as seen from Eq.~(\ref{time}) even a small change in $\epsilon_m$ can have a strong influence on the tunneling time when two conditions are fulfilled: (1) the temperature is small ($k_{\rm B}T < \Delta E$), so that the Fermi function $f$ has a steep slope near $\epsilon_m$ and (2) the degeneracies of the charging states $m$ and $m-1$ are different, so that $\xi_m \neq 1$.

For example, at the transition $m=1$ with $d_0=1$ and $d_1=2$, the relaxation rate is $\tau^{-1} = \Gamma [ 1+f(\epsilon_1) ]$.
At this point, we should emphasize again the importance of the charging energy.
For negligible charging energy, the charging and discharging of each quantum dot level (with orbital and spin degree of freedom) is independent of the other levels. 
Therefore, degeneracies would not play any role, $\xi=1$, and the relaxation time would be energy independent.
The finite asymmetry is, therefore, a clear signature of Coulomb interaction. 
A similar conclusion has been drawn from measurements of the width of tunneling resonances in quantum dots that are asymmetrically coupled to source and drain electrodes \cite{Koenemann}. 
There the dependence of the width on the polarity of the applied bias voltage could also be traced back to the energy dependence of $\Gamma[1+f(\epsilon)]$. 
On the other hand, identical relaxation times for charging and discharging have been recently observed on an electrostatically-defined quantum dot, coupled to a large top-gate capacitance (such that, the charging energy is negligible) \cite{Feve}. 

For an individual quantum dot or an ensemble with a sharp distribution of the quantum-dot resonances, asymmetric charge-relaxation times can only be observed when the final gate voltage after the voltage pulse for charging is {\it different} from the one for discharging, such that the corresponding quantum-dot level positions are separated by at least $k_{\rm B}T$. In our sample, however, the opposite limit of a rather broad distribution is realized.
To describe this case, we integrate over all energies for the quantum-dot levels in the ensemble and obtain an expression that is independent of the gate voltage after the pulse,
\begin{equation}
\label{occ}
  	\Delta N (t) \propto \int d \epsilon \left[ N_{\rm eq} (\epsilon\pm \Delta E) -  N_{\rm eq} (\epsilon)\right]  e^{- t/\tau(\epsilon)} \, .
\end{equation}
Here, $N_{\rm eq} (\epsilon\pm \Delta E)$ and $N_{\rm eq} (\epsilon)$ are the equilibrium occupation of the quantum dot before and after the voltage pulse, respectively.
The upper (lower) sign corresponds to charging (discharging).
The pre-exponential factor in the integrand selects only those quantum dots which change their occupation after the voltage pulse. 
An asymmetry of the relaxation times now appears because relative to the Fermi energy, the dot energies lie by $\Delta E$ lower for charging than for discharging. 
Due to the energy integral, the charge relaxation is no longer governed by a single exponential decay. 
To characterize the relaxation by a single time constant, we numerically perform an exponential fit of Eq.~(\ref{occ}). 

We quantify the asymmetry in the charge relaxation time by the ratio
\begin{equation}
	\label{ratio}
	\nu_m = \frac{\tau_{m\rightarrow m-1}}{\tau_{m-1\rightarrow m}} \, .
\end{equation}
For the reasons discussed above, $\nu_m$ is a function of temperature. It ranges from $\xi_m$ for $k_{\rm B}T \ll \Delta E$ to $1$ for $k_{\rm B}T \gg \Delta E$.

\section{Results and discussion}

Let us now consider the first two transitions $m=1$ and $m=2$ for filling the $s$-shell with the first and the second electron, respectively.
Spin degeneracy implies $d_0=1$, $d_1=2$ and $d_2=1$, which yields $\xi_1=2$ and $\xi_2=1/2$. 
Due to finite temperature, $\nu_1$ and $\nu_2$ should be closer to $1$ than $\xi_1$ and $\xi_2$.
Indeed, we measure $\nu_1= 1.4$ and $\nu_2=0.85$ at $T=4\,{\rm K}$.
Deviations from the expected relation $\nu_1=1/\nu_2$ may be attributed to an energy dependence of the tunneling barrier $\Gamma(\epsilon)$.
This effect can be accounted for by averaging $\nu_1$ and $1/\nu_2$ to $\bar\nu=1.3$.
A temperature-dependent comparison between measured and calculated values of $\bar\nu$ is shown in the inset of Fig.~\ref{Fig2}.
We find qualitative agreement: $\bar \nu$ ranges between $2$ and $1$ and decreases with temperature, where the crossover temperature is given by $k_{\rm B}T \sim \Delta E$.  
Quantitatively, the measured values of $\bar \nu$ are somewhat smaller than the calculated ones. 
A better agreement can be achieved by assuming a higher electron temperature, caused by Ohmic heating of the 2DEG through the measurement current and the voltage pulse. 
Also, fluctuations of the distance between quantum dots and 2DEG, i.e., variations of $\Gamma$ within the dot ensemble, may contribute to the discrepancy.

\begin{figure}[bt]
	\includegraphics[width=1.0\columnwidth]{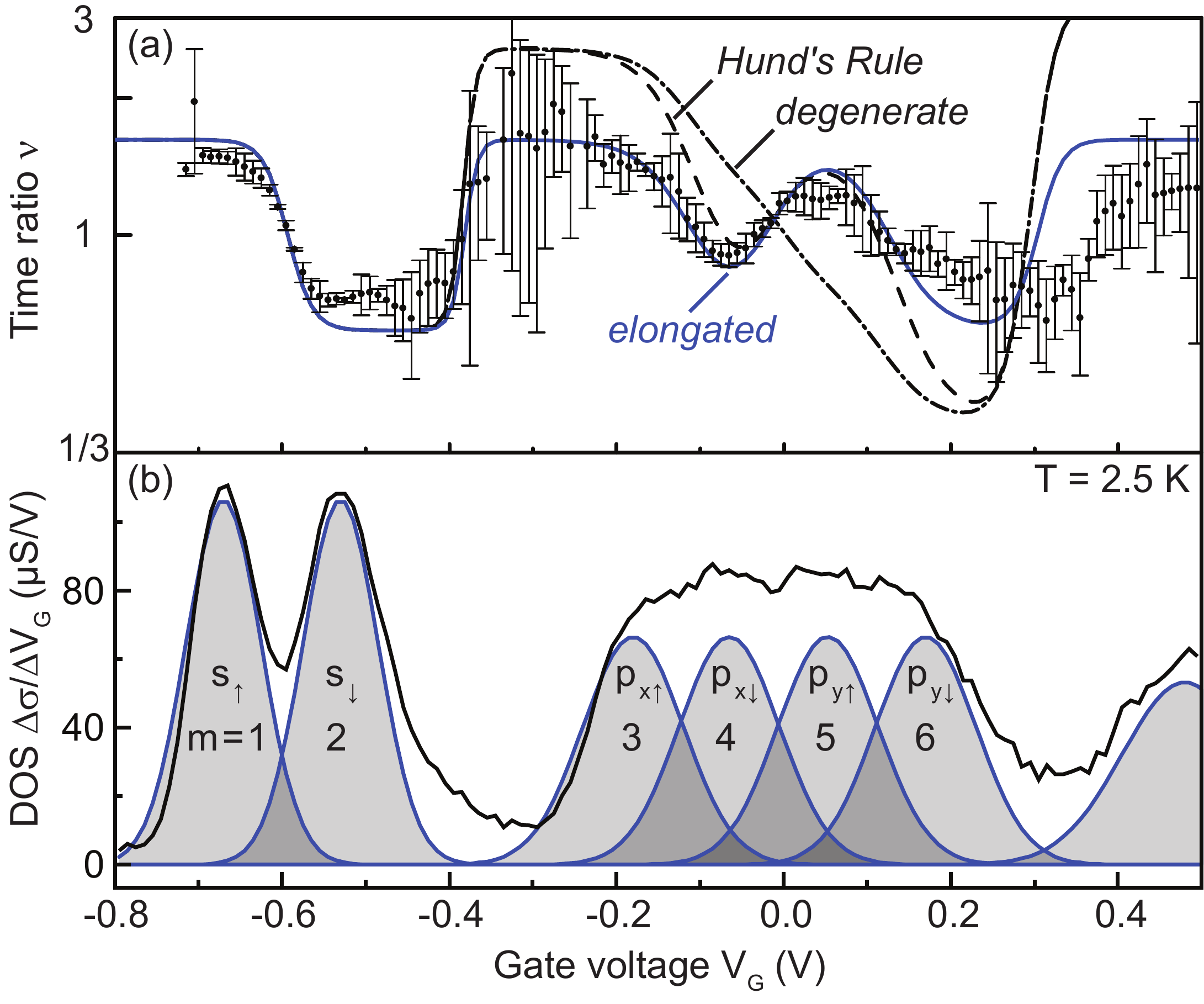}
	\caption{(a) Measured (dots) and calculated (lines) tunneling ratios using $\nu_m$ from table \ref{tab:nutable} for all transitions weighted with the average occupation of the ensemble. 
(b) Conductance change (i.e. density of states) due to charge transfer into the QDs. 
Shaded areas show fits to the measured density, which has been used for the calculation shown in (a).}
	\label{Fig3}
\end{figure}

So far, we have only looked at transitions involving the $s$-shell. 
Now, we turn to the filling sequence up to the sixth electron occupying the $p$-shell of the quantum dots. 
Even though for electrons, the $s, p, d\,\dots$ shell filling sequence has been verified repeatedly \cite{Bimbergbuch,Drexler94,Fricke96,Marquardt11,Cockins10}, it was not quite clear whether Hund's rule applies to the filling of the $p$-shell \cite{Tarucha96,Warburton98} or whether it is lifted by an anisotropy of the confinement potential in the dot \cite{Fricke96,Wibbelhoff05,Lei10}. 
Our time-resolved transconductance spectroscopy provides an excellent tool to clarify which scenario is realized. 
In Fig.~\ref{Fig3}(a), the data points show the experimental ratios $\nu$ as a function of gate voltage for a temperature of 2.5~K. To compare with our model, we consider three different scenarios: (i) a circular dot with non-interacting charge carriers, which gives degeneracies in the $p$-shell $\lbrace d_3, d_4, d_5, d_6\rbrace = \lbrace 4, 6, 4, 1\rbrace$, (ii) a circular dot, but taking Hund's rule into account, leading to degeneracies $\lbrace 4, 2, 4, 1\rbrace$, and (iii) an elongated dot with degeneracies $\lbrace 2, 1, 2, 1\rbrace$. 
The resulting $\xi_m$ and the calculated $\nu_m$ are listed in table \ref{tab:nutable}. 

Since the separation of the charging states in the $p$-shell is comparable to the inhomogeneous width of the QD ensemble, different transitions may occur during a single switch in energy. 
To account for this overlap, we use the decomposition of the density of states shown as shaded areas in Fig. \ref{Fig3}(b) to weight the processes with the percentage of dots at a certain occupation $m$. In Fig. \ref{Fig3}(a), the thus obtained  $\nu(V_{\rm G})$ are shown as dash-dotted, dashed, and solid lines for the models (i) -- (iii), respectively. 
Without any adjustable parameters, we find very good agreement with the model for an elongated dot, where all but the spin degeneracies have been lifted by the asymmetric potential. The other models are incompatible with the data. 
This finding is in agreement with wave-function mapping experiments \cite{Wibbelhoff05,Vdovin07,Beckel12}. 

It should be pointed out that the distribution of quantum-dot energy levels is much broader than the splitting $\delta E_p$ of the $p$-orbitals.
As a consequence, it is not possible to resolve $\delta E_p$ in the equilibrium density of states as shown in Fig.~\ref{Fig3}(b).
Nevertheless, from the time-resolved measurement, Fig.~\ref{Fig3}(a), we can unambiguously conclude that there is a splitting $\delta E_p$, which is larger than the energy shift $\Delta E$ caused by the voltage pulse.

Our calculations show that it may be possible to quantitatively determine an energy splitting $\delta E$ with our method, even when it is masked by inhomogeneous broadening:
For voltage pulses large enough such that $\Delta E \gtrsim \delta E$ one could map out the crossover from $k_{\rm B}T \gg \delta E$ for which the splitting can be neglected (larger degeneracy) to $k_{\rm B}T \ll \delta E$ for which the split levels are filled separately (smaller degeneracy). 
In the crossover regime, $k_{\rm B}T \sim \delta E$, charge and (for Zeeman splitting) spin dynamics are coupled to each other \cite{Splettstoesser,Contreras}.

\begin{table}[htb]
	\caption{Degeneracy ratios $\xi_m$ and relaxation time ratios $\nu_m$, calculated for different models of shell filling at $T=2.5$~K. Also shown are measured tunneling ratios $\nu$, determined by fits to the charging and discharging data (cf. Fig.~\ref{Fig2}) at the charging voltages of the $m^{\rm th}$ electron. Best agreement is found for the model of an elongated dot. }
\begin{tabular}{| r || c c | c c || c c | c c | c c | c c |}
\hline
Model	&	$\xi_1$	&	$\nu_1$	&	$\xi_2$	&	$\nu_2$	&	$\xi_3$	&	$\nu_3$	&	$\xi_4$	&	$\nu_4$	&	$\xi_5$	&	$\nu_5$	&	$\xi_6$	&	$\nu_6$	\\
\emph{degenerate}	&	2	&	1.6	&	$\frac{1}{2}$	&	0.6	&	4	&	2.6	&	$\frac{3}{2}$	&	1.3	&	$\frac{2}{3}$	&	0.8	&	$\frac{1}{4}$	&	0.4	\\
\emph{Hund's rule}	&	2	&	1.6	&	$\frac{1}{2}$	&	0.6	&	4	&	2.6	&	$\frac{1}{2}$	&	0.6	&	2	&	1.6	&	$\frac{1}{4}$	&	0.4	\\
\emph{elongated}	&	2	&	1.6	&	$\frac{1}{2}$	&	0.6	&	2	&	1.6	&	$\frac{1}{2}$	&	0.6	&	2	&	1.6	&	$\frac{1}{2}$	&	0.6	\\
\hline
measured &{}& 1.5 &{}& 0.7 &{}& 1.6 &{}& 0.7 &{}& 1.5 &{}& 0.7 \\
\hline
\end{tabular}
	\label{tab:nutable}
\end{table}

In conclusion, we propose time-resolved transconductance spectroscopy of quantum dots coupled to a 2DEG as useful tool to determine the degeneracies of the quantum-dot levels with a much better resolution than the inhomogeneous width of the QD ensemble. 
As a consequence of Coulomb interaction, the ratios of the charge relaxation times for charging and discharging is, in general, different from 1 and depends both on the level degeneracies and temperature. 
Our measurements can be qualitatively explained within a master-equation approach and they unambiguously show the existence of a $p$-orbital splitting. 

\section*{Acknowledgement}

We acknowledge financial support by the Mercator Research Center Ruhr (MERCUR) of Stiftung Mercator,
the DFG (Contract No. GE 2141/1-1) in the framework of the NanoSci-E+ project QD2D of the European Commission and the project QuaHL-Rep 16BQ1035 as well as the project 'Hochfunktionale Speicher' (HOFUS) within the VIP program of the BMBF.

\end{document}